\documentclass[aps,prd,preprint,superscriptaddress,nofootinbib,showpacs]{revtex4}
\usepackage{graphicx}

\usepackage{ulem}
\usepackage{color}
\definecolor{My_red}        {cmyk}{0.00,1.00,1.00,0.20}

\newcommand{\bmat}{\left(\begin{array}}
\newcommand{\emat}{\end{array}\right)}
\newcommand{\beq}{\begin{equation}}
\newcommand{\eeq}{\end{equation}}
\newcommand{\wt}{\widetilde}

\def\ra{\rightarrow}

\def\ld{\lambda}
\def\f{\frac}

\def\bwt{\begin{widetext}}
\def\ewt{\end{widetext}}
\def\be{\begin{equation}}
\def\ee{\end{equation}}
\def\bea{\begin{eqnarray}}
\def\eea{\end{eqnarray}}
\def\bean{\begin{eqnarray*}}
\def\eean{\end{eqnarray*}}
\def\bary{\begin{array}}
\def\eary{\end{array}}
\def\bit{\begin{itemize}}
\def\eit{\end{itemize}}

\def\ra{\rightarrow}

\def\ld{\lambda}

\def\su5u1{SU(5) \times U(1)}
\def\fsu5u1{SU(5) \times U(1)'}
\def\so10{SO(10)}
\def\sq20{SO(10) \times SO(10)}

\def\ra{\rightarrow}

\def\ld{\lambda}
\def\f{\frac}

\def\L{\left(}
\def\R{\right)}
\def\bwt{\begin{widetext}}
\def\ewt{\end{widetext}}
\def\be{\begin{equation}}
\def\ee{\end{equation}}
\def\bea{\begin{eqnarray}}
\def\eea{\end{eqnarray}}
\def\bean{\begin{eqnarray*}}
\def\eean{\end{eqnarray*}}
\def\bary{\begin{array}}
\def\eary{\end{array}}
\def\bit{\begin{itemize}}
\def\eit{\end{itemize}}

\def\ra{\rightarrow}

\def\ld{\lambda}

\def\su5u1{SU(5) \times U(1)}
\def\fsu5u1{SU(5) \times U(1)'}
\def\so10{SO(10)}
\def\sq20{SO(10) \times SO(10)}

\usepackage[centertags]{amsmath}
\usepackage{amssymb}

\begin{document}

\title{ The Maximal $U(1)_L$ Inverse Seesaw from  $d=5$ Operator and \\
  Oscillating Asymmetric Sneutrino Dark Matter}

\author{Zhaofeng Kang}
\email[E-mail: ]{zhaofengkang@gmail.com}
\affiliation{School of Physics, Korea Institute for Advanced Study,
Seoul 130-722, Korea}
\affiliation{Key Laboratory of Frontiers in Theoretical Physics,
      Institute of Theoretical Physics, Chinese Academy of Sciences,
Beijing 100190, P. R. China }

\author{Jinmian Li}
\email[E-mail: ]{phyljm@gmail.com}
\affiliation{ARC Centre of Excellence for Particle Physics at the Terascale, Department of Physics, University of Adelaide, Adelaide, SA 5005, Australia}
\affiliation{Key Laboratory of Frontiers in Theoretical Physics,
      Institute of Theoretical Physics, Chinese Academy of Sciences,
Beijing 100190, P. R. China }

\author{Tianjun Li}

\affiliation{Key Laboratory of Frontiers in Theoretical Physics,
      Institute of Theoretical Physics, Chinese Academy of Sciences,
Beijing 100190, P. R. China }

\affiliation{George P. and Cynthia W. Mitchell Institute for
Fundamental Physics, Texas A$\&$M University, College Station, TX
77843, USA }

\author{Tao Liu}

\affiliation{Key Laboratory of Frontiers in Theoretical Physics,
      Institute of Theoretical Physics, Chinese Academy of Sciences,
Beijing 100190, P. R. China }

\author{Jin Min Yang}

\affiliation{Key Laboratory of Frontiers in Theoretical Physics,
      Institute of Theoretical Physics, Chinese Academy of Sciences,
Beijing 100190, P. R. China }

\date{\today}

\begin{abstract}

The maximal $U(1)_L$ supersymmetric inverse seesaw mechanism (M$L$SIS) provides a natural way to relate asymmetric dark matter (ADM) with neutrino physics. In this paper we point out that, M$L$SIS is a natural outcome if one dynamically realizes the inverse seesaw mechanism in the next-to minimal supersymmetric standard model (NMSSM) via the dimension-five operator $(N)^2S^2/M_*$, with $S$ the NMSSM singlet developing TeV scale VEV; it slightly violates lepton number due to the suppression by the fundamental scale $M_*$, thus preserving $U(1)_L$ maximally. The resulting sneutrino is a distinguishable ADM candidate, oscillating and favored to have weak scale mass. A fairly large annihilating cross section of such a heavy ADM is available due to the presence of singlet.

\end{abstract}

\maketitle

\section{Introduction and Motivation}

Origins of tiny but not vanishing neutrino masses are of great interest. Among those, the inverse seesaw mechanism~\cite{Mohapatra:1986bd} gains special attention, mainly because it provides a natural, simple and testable way to realize small neutrino masses at low energy without invoking suppressed couplings. Besides, this mechanism follows the symmetry principle: a tiny neutrino mass, which slightly breaks lepton number by two units, is closely related to the degree of lepton number symmetry $U(1)_L$ violation. Such an observation yields a deep implication to the supersymmetric dark matter (DM) candidates in the supersymmetric standard models (SSMs): if the inverse seesaw mechanism is realized with retaining a maximal $U(1)_L$, i.e., one attributes the lightness of neutrino to $U(1)_L$ violation to the maximum extent (We refer to Eq.~(\ref{mass}) for a more detailed explanation.), the lightest sneutrino can be an asymmetric dark matter (ADM) candidate. And the resulting scenario is dubbed as M$L$SIS, the maximal $U(1)_L$ supersymmetric inverse seesaw.

Thus far, ADM~\cite{Barr:1990ca,Kaplan:2009ag,Kaplan:2009ag1,XO, Zurek:2013wia} is the most attractive mechanism to understand the coincidence between the relic densities of the dark and baryonic matters, $\Omega_{\rm DM}:\Omega_{b}\simeq 5:1$. But realizing the ADM scenario in SSMs usually requires a bulk of extension, for example, invoking higher dimensional operators with new scales~\cite{Kaplan:2009ag}. On the other hand, it was believed that the low scale supersymmetric type-I seesaw could provide sneutrino as an economic ADM candidate~\cite{Hooper:2004dc}, but it is rendered to be the ordinary symmetric DM by the large $U(1)_L$ violation effect~\cite{Chen:2015yuz}.~\footnote{The ordinary symmetric sneutrino dark matter is studied well by a lot of groups~\cite{sneutrino:DM}.} In contrast, in the M$L$SIS, by definition, the degree of $U(1)_L$ violation is under control: it just regenerates the symmetric DM components via oscillation~\cite{osci} but not spoils the ADM picture. The oscillating snuetrino ADM is strongly favored to have mass around the weak scale instead of the conventional GeV scale~\cite{Chen:2015yuz}.

Despite of being a well-motivated scenario to embed ADM in SSMs and moreover providing a distinguishable ADM candidate, the M$L$SIS, in the sense of model building, still can be improved from two aspects. First, the origin of the maximal lepton number, or the minimal $U(1)_L$ violation, is of concern. It is not a new problem but inherits from the inverse seesaw mechanism; see some attempts to address this problem~\cite{Bazzocchi:2009kc,Dias:2012xp}. As the central result of this paper, we find that the presence of a singlet $S$ developing TeV scale vacuum expectation value (VEV) provides a quite simple solution via a dimension-five operator with a high cut-off scale. Such a singlet is furnished in the well-known next-to minimal SSM (NMSSM)~\cite{Ellwanger:2009dp}, which thus provides the basis for model building. Second, in the minimally realized M$L$SIS~\cite{Chen:2015yuz} the sneutrino ADM fails in annihilating away effectively, and again a singlet can help us to cope with this problem.

This work is organized as follows. In Section II the M$L$SIS is realized in the $Z_3$-NMSSM with a dimension-five operator. In Section III, we study the oscillating sneutrino asymmetric DM, focusing on the annihilation. The conclusion is given in  Section IV. 

\section{The maximal $U(1)_L$  inverse seesaw based on NMSSM}

Let us begin with a brief review of the M$L$SIS, which is firstly proposed in Ref.~\cite{Chen:2015yuz}. In the minimal scenario, the superpotential is nothing but that of the supersymmetric inverse seesaw mechanism~\cite{Arina:2008bb,Guo:2013sna}:
\begin{align}\label{minimal}
W_{\rm IS}=y_NH_uLN^c+{m_N}NN^c+\f{M_{N}}{2}N^2.
\end{align}
We follow the notation of Ref.~\cite{Chen:2015yuz}: the chiral superfields are denoted as $N^c=(\wt \nu_R^*,\nu_R^\dagger)$ and $N=(\wt \nu_L',\nu_L')$, with $\nu_L'$ and $\nu_R$  both carrying lepton number $+1$. The Majorana mass term is the source of $U(1)_L$ violation, by two units. For simplicity, we consider the single family case. In the flavor basis $(\nu_L,\nu_R^\dagger,\nu_L')$, the neutrino mass matrix is given by
\begin{align}\label{}
M_{inverse}=\left(\begin{array}{ccc}
              0 & m_D & 0 \\
              m_D & 0 & m_N \\
              0 & m_N & M_{N}
            \end{array}\right),
\end{align}
with Dirac neutrino mass $m_D=y_N\langle H_u^0\rangle$. In order to avoid large non-unitarity, we impose the bound $K\equiv m_N/m_D\gtrsim 10$~\cite{nonU}. Then the lightest neutrino is dominated by the active neutrino:  $\nu_{1}\approx \sin{\theta_\nu} \nu_L-\cos{\theta_\nu} \nu_L'$ with $\sin{\theta_\nu}\approx 1-1/2K^2\approx 1$. The neutrino mass takes the form of double suppression
\begin{align}\label{mass}
m_{\nu}^{eff}=-\f{m_D^2}{m_N^2+m_D^2}M_{N}\simeq -M_N/K^2.
\end{align}
If $K$ takes a value as small as possible, $M_N$ should take the smallest value accordingly. So, $U(1)_L$ would be respected to the greatest extent, leading to the maximal $U(1)_L$.

The other two Weyl fermions  $\nu_{2,3} \approx \f{1}{\sqrt{2}}\L \pm  \nu_R^\dagger+ {\sin{\theta_\nu}}\nu'_L+\cos{\theta_\nu} \nu_L\R$ are singlet-like and heavy. They have almost degenerate masses $|M_{2,3}|=\sqrt{m_N^2+m_D^2}+{\cal O}(M_{N})\approx m_N$ and form a pseudo-Dirac fermion.

\subsection{ Realizing M$L$SIS in NMSSM via a dimension-five operator}\label{d=5}

In the M$L$SIS, $M_N$ is required to be $\lesssim10$ eV. Such a tiny mass scale implies that the $U(1)_L$ breaking term may originate from a higher dimension operator, which resembles the understanding on the active neutrino mass via the Weinberg operator ${\cal O}_{win}=(LH_u)^2/M^*$. Owing to the fact that the weak scale $v_u\simeq 246$ GeV is relatively low, to give the realistic neutrino mass we need a somewhat peculiar scale $M_*\sim 10^{14}$ GeV, which is close but two orders of magnitude lower than the grand unification theory (GUT) scale $\sim 10^{16}$ GeV. It is even far less than another putative fundamental scale, the Planck scale $M_{\rm Pl}\sim 10^{18}$ GeV or the string scale that interpolates between them.

In the case under consideration, the situation becomes quite different and intriguing new possibilities open. In order to construct a Weinberg operator-like operator for the $U(1)_L$ breaking mass term, a scalar singlet $S$ is introduced; moreover, it develops a VEV $v_s\equiv \langle S\rangle$ so that we have the analogy
\begin{align}\label{}
\f{(LH_u)(LH_u)}{{M_*}}\ra\f{N N SS}{M_*}.
\end{align}
Now we have $m_{\nu}^{eff}\simeq- \ld_2 v_s^2/(K^2M_*)$.
Given a multi-TeV $v_s$, $M_*$ can be naturally identified as the GUT scale for operator coefficient $\ld_2\sim1$. However, if $v_s$ is merely at the sub-TeV scale, we need to allow a large coefficient $\ld_2\sim K^2$. In particular, if we have $v_s\sim{\cal O}$(10) TeV, even $M_*=M_{\rm Pl}$ is possible. In this article we prefer a lower $v_s$ because then one can enjoy the benefits of NMSSM: enhancing the SM-like Higgs boson mass via the new quartic term $\ld^2|H_uH_d|^2$ without losing electroweak scale naturalness, i.e., keeping a smaller $\mu=\ld v_s\sim{\cal O}$(100) GeV~\cite{Ellwanger:2011aa,Kang:2012sy,Cao:2012fz}.

In SSMs, such a singlet is very welcome. As is well known, the minimal SSM (MSSM) contains an unique mass parameter in the superpotential, i.e., the $\mu$ parameter of the mass term for Higgsinos $\mu H_uH_d$. It is expected to be around the weak scale, which is technically natural but the origin of such a low scale should be addressed. Among others, the NMSSM provides a simple and attractive solution by 
updating $\mu$ to be a dynamic field, $\mu\ra S$~\cite{Ellwanger:2009dp}. As a bonus, $S$ can also generate the supersymmetric Dirac mass term for the singlets $N$ and $N^c$ in the M$L$SIS. So, we propose the following scale invariant  (or $Z_3-$invariant) superpotential except for the dimension-five operator:
\begin{align}\label{MDIS}
W=&W_{\rm NMSSM}+\L y_{N}L H_uN^c+{\ld_{1}}SN  N^c\R+\f{\ld_{2}}{4M_*}S^2N^2,\\
-{\cal L}^{soft}=&\left( m_{\wt L}|\wt L|^2+m_{\wt \nu_L^{\prime}}|\wt \nu_L^{\prime}|^2+m_{\wt \nu_R}|\wt \nu_R|^2\right)\cr
&+y_NA_NH_u\wt L\wt \nu_R^*+B_m{m_N}\wt \nu_L^{\prime}\wt \nu_R^*+\frac{B_{M}M_{N}}{2}(\wt \nu_L^{\prime})^2 + h.c.,
\label{MDIS:soft}
\end{align}

The soft SUSY-breaking parameters $A_N$, $A_1$, etc., are assumed to be real and around the weak scale.~\footnote{We do not introduce lepton flavor violating mass terms in the soft SUSY-breaking sector. Otherwise, the realization of the oscillating sneutrino ADM would be changed significantly.} The ordinary NMSSM sector with $Z_3$ symmetry takes the form of
\begin{align}\label{NMSSM}
W_{\rm NMSSM}=& \ld SH_u  H_d+\f{\kappa}{3}S^3,\\
-{\cal L}^{soft}_{\rm NMSSM}=&m_{H_u}^2|H_u|^2+m_{H_d}^2|H_d|^2+m_S^2|S|^2 +\L
\ld A_\ld SH_u   H_d+\f{\kappa}{3}A_\kappa S^3+h.c.\R.
\end{align}
As usual, we insist on the perturbative bound on the dimensionless couplings, e.g. $\ld\lesssim0.7$. After $S$ developing a VEV, all the mass terms in the superpotential Eq.~(\ref{minimal}) just like $\mu$ are dynamically generated,
\begin{eqnarray}\label{Majorana}
m_N=\ld_1 v_s,\quad 
M_{N}=\frac{\ld_2 v_s^2}{M_*}.
\end{eqnarray}
The simple model can provide all the elements we need and it is the minimal model to  dynamically realize the inverse seesaw mechanism because we do not introduce any new fields (only the NMSSM plus right-handed neutrinos).

We would like to stress that the inverse seesaw mechanism based on a dimension-five operator is bound to be realized maximally preserving $U(1)_L$. The reason is simple. From Eq.~(\ref{mass}) and  Eq.~(\ref{Majorana}) one obtains $v_s\sim K\sqrt{m_{\nu}^{eff}M_*/\ld_2}$. Thus, for the given $m_{\nu}^{eff} \sim 0.1$ eV, $M_*\sim M_{\rm GUT}$ and a not very large $\ld_2$, a large $K\gg10$ would push $v_s$ far above the TeV scale, hence losing the benefits of NMSSM stated before. Therefore, we want the $K$ to be as small as possible, giving rise to the M$L$SIS scenario.

\subsection{Tentative  UV completion}

Since the small neutrino mass scale is simply a relic of fundamental scale physics, this inspires us to investigate the possible models at the fundamental scale. We find that a new $U(1)_R'$ symmetry can guarantee the general form of our model. At the renormalizable level, the supersymmetric model described by Eq.~(\ref{NMSSM}) plus
Eq.~(\ref{MDIS}) possesses an accidental $U(1)_B\times U(1)_L\times U(1)_R$ symmetry with the field charges assigned as
\begin{eqnarray}\label{assi}
&&L: \quad H_u[0],\,\,H_d[0],\,\,S[0],\,\,L[1],\,\,E^c[-1],\,\,N^c[-1],\,\,N[1],\nonumber\\
&& B: \quad H_u[0],\,\,H_d[0],\,\,S[0],\,\,Q[1],\,\,U^c[-1],\,\,D^c[-1],\,\,N[0],\nonumber\\
&& R: \quad H_u[2/3],\,\,H_d[2/3],\,\,S[2/3],\,\,L[2/3],\,\,E^c[2/3], \cdots \nonumber\\
&& R': \quad H_u[2/3],\,\,H_d[2/3],\,\,S[2/3],\,\,L[1/3],\,\,E^c[4/3], \,N^c[4/3], \,N[1/3],\,\Phi[1]\cdots
\end{eqnarray}
where the dots denote all other fields carrying the same charge $2/3$. As a matter of fact, the $U(1)_R$ charge assignment is not fixed according to this superpotential and in the above we simply choose one as an example,
which is consistent with $SU(5)$ GUT. Note that the $Z_3$ symmetry simply is an accidental result of $U(1)_R$ symmetry, which forbids the bare mass terms. At the dimension-five level, the operator $S^2 N^2$ violates the global symmetry $U(1)_L$ and $U(1)_R$ simultaneously, but still leaves a discrete $Z_2^L\subset U(1)_L$
and a new $U(1)_R'$ invariance, $R'\equiv R-\f{1}{3}L$. The $R'$ charge assignment of various fields is presented in the last line of Eq.~(\ref{assi}). In particular, if $U(1)_R'$ is generated to all orders, it was found that as a consequence $U(1)_B$ and the matter parity $Z_2^{M}\equiv (-1)^{3(B-L)}$ are conserved to all orders \cite{Lazarides:1998iq}.

With such a $U(1)'_R$ symmetry, we try to explore concrete UV completions of the low energy model which contains dimension-five operator and thus hints for new physics. This is of concern, since we will find that $M_*$ tends to be far below the fundamental Planck scale. We introduce a heavy singlet $\Phi$ carrying unit $U(1)'_R$ charge and thus it can (only) couple to $S$ and $N$ via the renormalizable term:
\begin{align}\label{}
W=\ld_\Phi SN\Phi + \f{M_\Phi}{2}\Phi^2,
\end{align}
with $M_\Phi\sim M_*$. Now integrating out $\Phi$ via the $F-$flatness condition of $\Phi$, namely $F_\Phi=M_\Phi\Phi+\ld_\Phi SN=0$, one then obtains the operator $\f{1}{M_*}S^2N^2$ with $M_*=- M_\Phi/2\ld_\Phi^2$. We would like to point out that, in the presence of three families of RHNs, one may arrange an accident hierarchy among $\ld_\Phi$ or (and) $M_\Phi$ such that one effective cutoff scale $M_*$ is hierarchically larger than others, and consequently the corresponding $M_N$ is much smaller than others. Later, we will see that It is helpful to realize the oscillating sneutrino ADM.


\section{Oscillating asymmetric sneutrino dark matter}

\begin{figure}[htb]
\includegraphics[width=1.0\textwidth]{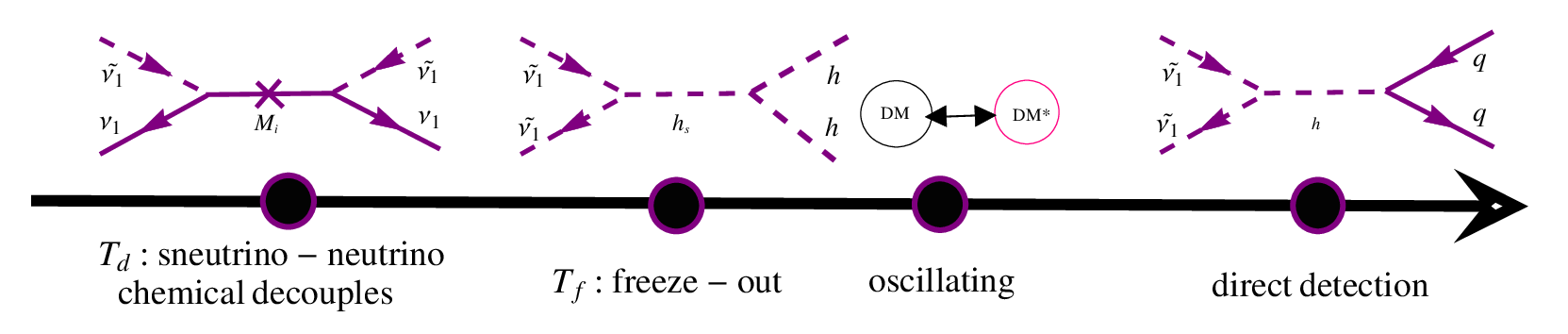}
\caption{Thermal history and dynamics of the oscillating sneutrino dark matter.}
\label{thermal} 
\end{figure}

In this section we will study the main phenomenology of M$L$SIS implemented in the NMSSM, oscillating asymmetric sneutrino dark matter. Although the main physics has been investigated in Ref.~\cite{Chen:2015yuz}, there are still several difference between the M$L$SIS with and without the singlet $S$; they will be the focuses here. We briefly discuss the similarities like asymmetry transfer and symmetry regeneration in the first subsection;  for more details, see Ref.~\cite{Chen:2015yuz}. And for illustration, we show the thermal history and the corresponding dynamics of sneutrino ADM in Fig.~\ref{thermal}.

\subsection{Profiles of the oscillating sneutrino ADM}\label{DM}

A big bonus of maximal $U(1)_L$ is the presence of an ADM candidate, the sneutrino. Let us begin with an exact $U(1)_L$ thus strictly complex sneutrinos. In the basis $\Phi^T=(\wt \nu_{L},\wt \nu_R,\wt \nu_L')^T$, the sneutrino mass squared matrix  is given by
\begin{align}\label{mL2}
            m_{\wt \nu}^2\approx&\left(\begin{array}{ccc}
              m_{\wt L}^2+m_D^2 &
\left( -m_D A_N+\mu m_D\cot\beta\right) & -m_D m_N \\
                & m_{\wt \nu_R}^2+m_N^2+m_D^2 & \f{1}{\sqrt{2}}A_1m_N \\
               &  & m_{\wt\nu_L'}^2+m_N^2 
            \end{array}\right).
\end{align}
Among three sneutrino $\wt\nu_{1,2,3}$ in the mass eigenstate, the lightest sneutrino is denoted as $\wt\nu_1$. The stringent DM direct detection requires that the left-handed sneutrino composition in $\wt\nu_1$ should be very small, and hence we can make the approximation:
\begin{align}\label{rotation}
\wt \nu_L'\approx-\sin\wt\theta \wt \nu_1+ \cos\wt\theta \wt \nu_2,\quad \wt \nu_R\approx\cos\wt\theta \wt \nu_1+ \sin\wt\theta \wt \nu_2,
\end{align}
with $\wt\theta$ the mixing angle. $\wt \nu_1$ gains asymmetry when it enters chemical equilibrium with the leptons; after the equilibrium breaks at $T_d$, the left asymmetry in ADM is $\eta_{0}\sim f_{\rm ADM}(x_d)\eta_b$, with $\eta_b\approx 10^{-11}$ the asymmetry of baryon. $f_{\rm ADM}(x_d)$ is a factor encoding the thermal threshold effect; it tends to be 1 in the relativistic limit $x_d=m_{\rm ADM}/T_d\ll 1$; in the opposite it is exponentially suppressed.

The above conventional picture of ADM may be spoiled by the tiny $U(1)_L$ violation. It induces mixing between the CP-even and -odd components of $\wt\nu_1=\f{1}{\sqrt{2}}\L{\rm Re}\wt\nu_1+I {\rm Im}\wt\nu_1\R$ and moreover splits their masses by an amount $\delta m$. Consequently, DM and anti-DM can oscillate into each other. If oscillation is significant during ADM freeze-out, ADM will turn out to be an ordinary symmetric DM. The oscillating rate is very sensitive to $\delta m$, whose upper limit is very sensitive to the ADM mass~\cite{oscillating,Tulin:2012re}: ADM $\sim 300$ GeV can tolerate $\delta m\sim 10^{-5}$ eV; but for the conventional GeV ADM, $\delta m$ is forced to be incredibly small, $\lesssim10^{-10}$ eV. So we will consider a weak scale sneutrino ADM.

However, even $\delta m\sim 10^{-5}$ eV is still hard to achieve in the M$L$SIS. To see this, one can well approximate the mass splitting as
\begin{align}\label{deltam}
\delta m\approx \f{\delta m_{11}^2}{m_{\wt\nu_1}}= \f{m_NM_{N}\sin2\wt\theta-B_{M}M_N\sin^2\wt\theta}{m_{\wt\nu_1}}.
\end{align}
As one can see, the natural scale of $\delta m$ should be not far below $M_N$ except for $\sin\wt\theta\ll 1$. However, for $m^{eff}_\nu\sim0.1$ eV one has $M_N\sim K^2 m_{\nu}^{eff}\sim {\rm eV}\gg 10^{-5}$ eV. Therefore, it is likely that $m^{eff}_\nu$ should be relaxed, says having a value $\ll 0.1$ eV. This is allowed in the three families of RHNs and may be regarded as a prediction of M$L$SIS with sneutrino ADM.

\subsection{Constraining $\ld_1$ from charge washing-out}\label{DM}

It is already noticed that a viable sneutrino ADM in the M$L$SIS needs the aid of a singlet to annihilate away the symmetric part through the term $\ld_1NN^c$~\cite{Chen:2015yuz}. But the magnitude of $\ld_1$ is stringently constrained by the DM charge violating scattering (CVS) process $\wt\nu_1\nu_1\leftrightarrow \wt\nu_1^*\bar\nu_1$, which is mediated by neutralinos and can keep chemical equilibrium between ADM and the light neutrinos until a quite low temperature $T_d$~\cite{Kang:2011ny}. If $T_d$ is down to the DM freeze-out temperature $T_f\sim m_{\rm DM}/20$, no asymmetry will be left.


To determine $T_d$, one has to compare the Hubble expansion rate $H(T)\approx 5.5T^2/M_{\rm Pl}$ with the CVS reaction rate, which can be obtained from the following effective Lagrangian: 
\begin{align}\label{}
-{\cal L}_{wash}=\f{1}{2}M_i^2\bar \chi_i\chi_i+\L y_{i1}\wt \nu_1^*\bar \chi_i P_L\nu_1+h.c.\R,
\end{align}
with $\chi_i$ the five Majorana neutralinos in the NMSSM. They are related to the states in the interacting eigenstates via $\chi_i=Z^T_{ij}\psi_j$ with $\psi=(\wt B,\wt W^3,\wt H_d^0,\wt H_u^0,\wt s)^T$. Approximately, the effective couplings $y_{i1}$ are given by  
\begin{align}\label{}
y_{i1}\approx y_N \sin{\theta_\nu}\cos\wt\theta Z_{4i}-\ld_1\cos{\theta_\nu} \cos\wt\theta Z_{5i},
\end{align} 
where the second is from the $\ld_1$-term. The CVS rate is calculated to be
\begin{align}\label{}
\Gamma_{\rm CVS}=5\times10^3\times\f{ |y_{i1}^2|^2}{12\pi^3}\L\f{T}{M_i}\R^4\L\f{T}{m_{\wt \nu_1}}\R^2 T.
\end{align}
Now, the condition $\Gamma_{\rm CVS}(T_d)<H(T_d)$ gives the upper bound on couplings
\begin{align}\label{upper}
{ |y_{i1}^2|^2}\lesssim0.41x_d \L\f{M_i}{T_d}\R^4 \f{m_{\wt \nu_1}}{M_{\rm Pl}}=1.0\times10^{-10}\L \f{M_i/m_{\wt \nu_1}}{10}\R^4\L\f{x_d}{5}\R^5\L\f{m_{\wt \nu_1}}{100\rm GeV}\R.
\end{align}
In the above estimation, ADM for reasons introduced later is assumed to be relatively heavy, around the weak scale, 
but neutralinos are even much heavier, having multi-TeV masses so as to suppress the CVS rate. Then, it is seen that $y_{i1}\lesssim10^{-2}$ should be fulfilled. But we typically need $y_{i1}\lesssim10^{-3}$ if neutralinos merely have masses close to the ADM mass, and it is probably true at least for Higgsinos, whose masses are mainly determined by the $\mu-$term, expected to lie around the weak scale for the sake of weak scale naturalness.

Now we investigate possible ways to get small $|y_{i1}|^2$ and the difficulty therein. First, neutrinos in the decoupling limit, i.e., $\cos\theta_\nu\approx1/K\ll1$, helps to suppress the $\ld_1-$contribution. However, we know that by definition M$L$SIS needs $K$ to be as small as possible, so we merely have $\cos\theta_\nu\sim 0.1$. Second, as long as $\wt\nu_1$ is dominated by $\wt{\nu}_L'$, all these couplings can be naturally small due to the suppression from $\cos\wt\theta\ll1$. But such a situation will hamper the attempt to decrease the mass splitting $\delta m$ (see Eq.~(\ref{deltam})). Of course, the smallness of $\delta m$ can always be attributed to a small $M_N$, so the option $\wt\nu_1\simeq \wt\nu_L'$ services as the last trick for avoiding large CVS.

\subsection{Annihilating away the symmetric part}

Now we are at the position to discuss the sneutrino ADM symmetric annihilation. The interactions between sneutrinos and the NMSSM sector heavily rely on $\ld_1$ and as well as $m_N$; see Eq.~(\ref{DM-Higgs}). We list the relevant terms for convenience:
\begin{align}\label{DM-Higgs1}
{\cal L}_{\wt\nu_1} 
\supset &
-i \L \f{\ld_1}{\sqrt{2}}A_1  -\sqrt{2}\kappa m_N  \R a_s \wt\nu_1^*\wt\nu_2 +\cos2\wt\theta  \L \f{\ld_1}{\sqrt{2}} A_1+\sqrt{2}\kappa m_N  \R  h_s \wt\nu_1^*\wt\nu_2 
\cr 
&
+\L\ld_1^2+\ld_1\kappa \sin2\wt\theta \R \f{a_s^2}{2}| \wt\nu_1|^2+ \L\ld_1^2-\ld_1\kappa \sin2\wt\theta \R \f{h_s^2}{2}| \wt\nu_1|^2
\cr 
&+ \left[ \sqrt{2}\ld_1 m_N -\f{\sin2\wt\theta}{\sqrt{2}}\L \ld_1A_1
 +2\kappa m_N\R\right] h_s | \wt\nu_1|^2-\f{\sin2\wt\theta}{\sqrt{2}}\ld\ld_1\L v_u h_d+v_dh_u\R   | \wt\nu_1|^2.&
\end{align}
Interactions involving $y_N=m_N/Kv_u\ll 1$ (to satisfy the CVS bound) are neglected. One may wonder if it is possible to get a large ADM annihilation cross section in the $\wt\theta\ra0$ limit ($\wt\nu_1\simeq \wt\nu_R$), which is favored by small $\delta m$. Unfortunately, we cannot. In that limit, the CVS bound requires $\ld_1\lesssim{\cal O}(0.01)$ and thus all of the couplings in Eq.~(\ref{DM-Higgs1}) are suppressed except for the massive coupling $\kappa m_N$, which may be sizable due to a large $m_N$. But it renders a large $y_N$, inconsistent with the CVS bound. In what follows we will present a viable scenario, characterized by a large $\ld_1\sim{\cal O}(0.1)$ and small $v_s$ at the sub-TeV scale.




Two ways are available to annihilate away the symmetric part with a cross section at least a few pb~\cite{Graesser}. One is annihilating into the lighter $a_s/h_s$~\footnote{Mixings are neglected in our estimation.} pair via the contact interactions, with cross sections $\simeq {\ld_1^4}/({64\pi} {m_{\wt\nu_1}^2})$. Thus it works for $\ld_1\simeq0.3$ and a lighter ADM with mass $m_{\wt\nu_1}\lesssim100$ GeV. The other one is via a $s$-channel $h_s$. Near the resonant enhancement region, the inclusive cross section is
\begin{align}\label{}
\sigma v= 4\pi \f{\Gamma(h_s\ra \wt\nu_1\wt\nu_1^*)\Gamma_{h_s}}{(s-m_{h_s}^2)^2+m_{h_s}^2\Gamma_{h_s}^2} \lesssim \f{\pi}{m^2_{\wt \nu_1}}\left[1-{\rm Br}(h_s\ra \wt\nu_1\wt\nu_1^*)\right].
\end{align} 
Hence in principle it can easily reach ${\cal O}$(pb) as long as $h_s$ does not dominantly decay into a pair of DM. Actually, $h_s$, due to a sizable $\ld$ near 1, tends to dominantly decay into a pair of SM-like Higgs bosons or Higgsinos if kinematically accessible.



\subsection{On the detections of sneutrino DM}

Sneutrnino DM can interact with quarks via the three Higgs bosons $H_i$, but the interaction strength are supposed to be fairly weak. One can see this from the last line in Eq.~(\ref{DM-Higgs1}), where $\sin2\wt\theta\ll 1$ in order to satisfy the CVS bound and thus the only sizable contribution is from the $\ld_1m_N-$term; moreover, this term is negligible unless $h_s$ strongly mixes with the doublet component. We consider this case to see the prospect of direct detection of ADM.


$H_i$ mediate DM-nucleon spin-independent (SI) scattering. Its cross section, normalized to DM-proton scattering, is conventionally written as $\sigma_{\rm SI}=4a_p^2\mu^2_p/\pi$ with $\mu_p$ the proton-DM reduced mass. The effective proton-DM coupling $a_p$ receives three contributions 
\begin{align}
a_{p,H_i}=\f{\mu_{H_i 11}}{2m_{\wt\nu_1}}\f{1}{m_{H_i}^2}\f{m_p}{v}\left[\sum_{q=u,d,s}f^{(p)}_{T_q}{g_{qqH_i}}
+\f{2}{27}\sum_{q=c,b,t}f^{(p)}_{T_G}{g_{qqH_i}}\right]^2,
\end{align}
where $\mu_{H_i 11}$ are the massive couplings for $H_i| \wt\nu_1|^2$; concretely, $\mu_{H_i 11}\approx \ld_1 m_N O_{i3}$. The effective couplings are $g_{uuH_i}=O_{i2}/\sin\beta$ for the up-type quarks and $g_{ddH_i}=O_{i1}/\cos\beta$ for the down type quarks with $O$ defined in Eq.~(\ref{O:de}). The coefficients take values $f_{T_u}^{(p)}=0.023, f_{T_d}^{(p)}=0.033, f_{T_s}^{(p)}=0.26$ and $f_{T_G}^{(p)}=1-\sum_{q=u,d,s}f^{(p)}_{T_q}=0.684$~\cite{Belanger,Gao:2011ka}. With them one can parameterize $a_{p,H_i}$ as
\begin{align}\label{api}
a_{p,H_i}=4.0\times10^{-3}\times\f{\mu_{H_i 11}}{2m_{\wt\nu_1}}\f{1}{m_{H_i}^2}\L 0.123\f{O_{i2}}{\sin\beta}+0.343\f{O_{i1}}{\cos\beta}\R.
\end{align}
For DM around the weak scale like 300 GeV, currently the most stringent upper bound $\sigma^{up}$ is from LUX~\cite{LUX}, about $10^{-9}$pb, implying $a_{p,H_i}\lesssim 1.6\times 10^{-9}\L \sigma^{up}/10^{-9}{\rm pb}\R^{1/2}\rm GeV^{-2}$. Typically, $\sigma_{\rm SI}$ here lies below the upper bound:
\begin{align}\label{}
a_{p,H_1}\approx0.8\times 10^{-9} \L\f{\ld_1m_N}{10\rm GeV}\R\L\f{200\rm GeV}{m_{\wt\nu_1}}\R \L\f{125\rm GeV}{m_{H_1}}\R^2\L\f{0.03}{\rm mixing}\R\rm GeV^{-2}.
\end{align}
In this optimistic estimation, $H_1$ is the SM-like Higgs boson and ``mixing" denotes the factor in the bracket of Eq.~(\ref{api}). But the next round of detection may reach the sneteutrino ADM. Of course, the most promising probe is from indirect detection, because our ADM possesses a large annihilation cross section today; it is totally different to the most ADM scenario except for the decaying one~\cite{Feng:2013vva}.

\section{Conclusion}

The M$L$SIS provides an attractive way to relate ADM with neutrino physics. Such a scenario is a necessary outcome if one dynamically realizes the inverse seesaw mechanism in the NMSSM via the dimension-five operator $(N)^2S^2/M_*$ to explain the origin of the smallness of lepton number violation. The sneutrino is a distinguishable ADM candidate, oscillating and favored to have weak scale mass. A fairly large annihilating cross section of such a heavy ADM is available due to the presence of singlet.

\section*{Acknowledgement}

This work was supported
by the National Natural Science Foundation of China under grant Nos.
10821504, 10725526 and 10635030, by the DOE grant
DE-FG03-95-Er-40917, and by the Mitchell-Heep Chair in High Energy
Physics.

\appendix

\section{Relevant interactions of sneutrino DM with Higgs bosons}

In studying the sneutrino DM annihilation and as well its scattering with nucleon, the interactions with Higgs bosons are relevant. We collect the dominant terms from $F-$term and the soft terms below
\begin{align}\label{DM-Higgs}
{\cal L}_{\wt\nu_1}\supset& |\kappa S^2+\ld H_uH_d+\ld_1 \wt \nu_L'\wt \nu_R^*|^2+|\ld_1S\wt\nu_R^*|^2+|\ld_1S\wt\nu_L'|^2\cr 
\supset& 
-i \L \f{\ld_1}{\sqrt{2}}A_1  -\sqrt{2}\kappa m_N  \R a_s \wt\nu_1^*\wt\nu_2 +\cos2\wt\theta  \L \f{\ld_1}{\sqrt{2}} A_1+\sqrt{2}\kappa m_N  \R  h_s \wt\nu_1^*\wt\nu_2 
\cr 
&
+i\f{\ld_1\ld}{\sqrt{2}}\L v_da_u+ v_ua_d\R  \wt\nu_1^*\wt\nu_2 + \f{\ld_1\ld}{\sqrt{2}}\cos2\wt\theta\L v_dh_u+ v_uh_d\R  \wt\nu_1^*\wt\nu_2 +c.c.
\cr 
&
+\L\ld_1^2+\ld_1\kappa \sin2\wt\theta \R \f{a_s^2}{2}| \wt\nu_1|^2+ \L\ld_1^2-\ld_1\kappa \sin2\wt\theta \R \f{h_s^2}{2}| \wt\nu_1|^2
-\ld\ld_1\f{\sin2\wt\theta}{2} \L h_u h_d-a_ua_d\R | \wt\nu_1|^2
\cr 
&
+  \left[ \sqrt{2}\ld_1 m_N -\f{\sin2\wt\theta}{\sqrt{2}}\L \ld_1A_1
 +2\kappa m_N\R\right] h_s | \wt\nu_1|^2-\f{\sin2\wt\theta}{\sqrt{2}}\ld\ld_1\L v_u h_d+v_dh_u\R   | \wt\nu_1|^2.
\end{align}
We have written the Higg fields as $S=v_s+\L h_s+ia_s\R/\sqrt{2}$ and similar to others.

We have not transformed the Higgs fields into their mass eigenstates yet. Following the convention in Ref.~\cite{Ellwanger:2009dp} we use matrix $O$ to do this for the CP-even Higgs bosons: 
\begin{eqnarray}\label{O:de}
(H_1,H_2,H_3)^T=O(h_d,h_u,h_s)^T,
\end{eqnarray}
with $H_i$ ordered in mass. As for the CP-odd Higgs bosons, we first work in the basis $(A,a_s)$ with $A=\cos\beta a_u+\sin\beta a_d$; the Goldstone mode $G=-\cos\beta a_d+\sin\beta a_u$ is projected out. Then we diagonalize $(A,a_s)$ using matrix $P'$: $(A_1,A_2)^T=P'(A,a_s)^T$. Finally we have
 \begin{eqnarray}
a_d=P_{i1}A_i,\quad a_u=P_{i2}A_i,\quad a_s=P_{i3}A_i,
\end{eqnarray}
with $P_{i1}=\sin\beta P_{i1}'$, $P_{i2}=\cos\beta P_{i1}'$ and $P_{i3}= P_{i2}'$ ($i=1,2$).





\end{document}